# Understanding Students' Acceptance of ChatGPT as a Translation Tool: A UTAUT Model Analysis

Lulu Wang[1]    Simin Xu[1]    Kanglong Liu[1,2]*

1 Department of Chinese and Bilingual Studies, The Hong Kong Polytechnic University

2 The Centre for Translation Studies，The Hong Kong Polytechnic University

**Author notes**

Correspondence should be addressed to Kanglong Liu, Department of Chinese and Bilingual Studies, The Hong Kong Polytechnic University, Hung Hom, Kowloon, Hong Kong SAR; Email: kl.liu@polyu.edu.hk

**Abstract**

The potential of ChatGPT to transform the education landscape is drawing increasing attention. With its translation-related capabilities being tested and examined, ChatGPT presents both opportunities and challenges for translation training. The effective integration of ChatGPT into translation training necessitates an understanding of students' reactions to and acceptance of ChatGPT-assisted translation. Against this backdrop, this study draws on the Unified Theory of Acceptance and Use of Technology (UTAUT) to examine the potential determinants of students' adoption of ChatGPT for translation and investigates the moderating effects of use experience and translation training on those relationships. An online survey targeting university students in Hong Kong collected 308 valid responses, including 148 from translation students and 160 from non-translation students. Respondents were divided into two groups based on their ChatGPT use experience. Data were analyzed using structural equation modeling. A multigroup analysis revealed different structural relationships between the influencing factors of students' intention to use ChatGPT across groups. Notably, less-experienced users' behavioral intention to use ChatGPT for translation was more strongly correlated with social influence compared with experienced users. Non-translation students' use intention was more strongly driven by facilitating conditions compared to translation majors. These results are discussed with the different primary purposes of translation and non-translation students' translation practices. The findings of this study contribute to the growing body of research on AI-powered translation training and provide insights for the ongoing adaptation of translation training programmes.

# 1. Introduction

Technological advancements have transformed translation practices in recent decades (Kenny, 2020). Integrating technological aids into translators' toolkits and translation curricula has long been acknowledged as necessary across industry and academia (Bowker, 2002; González Pastor, 2021; O'Brien, 2012; O'Hagan & Minako, 2013; Pym, 2013). Students unable to exploit translation technologies may be at a disadvantage in the translation market (Kenny, 2020). Thus, fostering students' technology literacy is a practical necessity in translation training. ChatGPT, a generative AI chatbot released by OpenAI in November 2022, has brought new opportunities for students and translation education (Farrokhnia, Banihashem, Noroozi, & Wals, 2023; Muñoz-Basols, Neville, Lafford, & Godev, 2023). As a large language model chatbot, ChatGPT can generate human-like responses and has shown potential as a powerful translation tool (Calvo-Ferrer, 2023; Hendy et al., 2023; Lee, 2023; Son & Kim, 2023).

Though not designed specifically for language translation, ChatGPT has demonstrated performance comparable to, or in some cases, superior to state-of-the-art neural machine translation systems like Google Translate, producing highly consistent, coherent, and precise translations (Lee, 2023). Some research even indicates that its output quality has achieved human parity for certain translation tasks (Calvo-Ferrer, 2023). Despite its impressive capabilities (Kalla & Smith, 2023), the use of ChatGPT in translation raises several concerns, notably issues of information privacy and output bias (Azaria, Azoulay, & Reches, 2023; O'Brien, 2023). Additionally, its translation outputs for low-resource languages, which have less available training data, are less satisfactory (Hendy et al., 2023; Jiao, Wang, Huang, Wang, & Tu, 2023; Son & Kim, 2023). Considering these complexities, integrating ChatGPT into translation training demands caution. Students need to be informed of its capabilities and limitations to use it wisely and effectively.

To effectively train students in using ChatGPT for translation, it is crucial to investigate their perceptions of ChatGPT-assisted translation. Such empirical insights are important for designing relevant courses (Austermuehl, 2013). However, research in this area is lacking, and most studies on students' acceptance of ChatGPT have only addressed its usage for general academic purposes, without focusing on a particular context (e.g., Liu & Ma, 2023; Romero-Rodríguez, Ramírez-Montoya, Buenestado-Fernández, & Lara-Lara, 2023; Strzelecki, 2023; Strzelecki & ElArabawy, 2024). As users' perceptions of technology can vary depending on the intended use (Lee, Kozar, & Larsen, 2003), the lack of focus in these studies limits the depth of insights that can be derived from them. The applicability of these findings to the field of translation remains uncertain. It is unclear to what extent students are willing to adopt ChatGPT for translation purposes and what factors impact their decision, especially considering the number of freely available machine translation systems that students are already familiar with (Carré, Kenny, Rossia, Sánchez-Gijón, & Torres-Hostench, 2022). Furthermore, the potential moderating factors for students' acceptance of ChatGPT in the context of translation are underexplored.

This study aims to investigate the factors influencing students' intention to use ChatGPT as a translation tool, drawing on the Unified Theory of Acceptance and Use of Technology (UTAUT) (Venkatesh, Morris, Davis, & Davis, 2003). Additionally, it examines the moderating effects of students' experience with ChatGPT and explores whether these factors exhibit the same patterns between translation and non-translation students. The research is guided by the following questions: 1) What factors influence students' intention to use ChatGPT as a translation tool? 2) How do students' use experience and their specialization affect their intention to use ChatGPT as a translation tool? By addressing these questions, this study aims to offer fresh and context-specific perspectives on students' use of AI tools. The findings will inform translation educators and AI system developers, ultimately enhancing students' learning experiences and outcomes.

## 2. Literature Review
### 2.1. ChatGPT as a Translation Tool and Implications for Students

ChatGPT has gained considerable attention in academia for its extraordinary language competence (Azaria et al., 2023). Several studies have specifically examined its machine translation capabilities, using both human evaluation and automatic metrics (Calvo-Ferrer, 2023; Lee, 2023; Son & Kim, 2023), providing preliminary insights into ChatGPT's strengths, limitations, and the potential benefits and risks associated with its use for translation tasks by students.

Calvo-Ferrer (2023) evaluated the quality of English-to-Spanish subtitles produced by ChatGPT and a human translator by having translation students rate their quality and distinguish between the two. Interestingly, the study found no significant difference in the quality scores assigned to the two types of subtitles, suggesting comparable quality between ChatGPT-generated and human translations. Lee (2023) investigated the Japanese-English language pair and analyzed the translation outputs of ChatGPT and three state-of-the-art machine translation systems: Google Translate, DeepL, and Microsoft Translator. According to Lee's analysis, ChatGPT outperforms the other systems with its more consistent and correct pronominal choices, precise wording, creative approach, and better coherence in its outputs. Son and Kim (2023) compared ChatGPT's translation quality with that of Google Translate and Microsoft Translator. The study involved eleven language pairs and both translation directions, using automatic translation quality evaluation systems. Despite the limitations of these metrics, Son and Kim reported that ChatGPT's translation quality falls behind the two state-of-the-art neural machine translation systems. Interestingly, ChatGPT performs better when translating into English than from English. Similarly, Hendy et al. (2023) conducted a large-scale comparison with other machine translation systems and suggested that ChatGPT's translation quality depends on the amount of available training data for a particular language, showing a performance gap between high-resource and low-resource languages.

In addition to the quality of translation outputs, ChatGPT offers features that may give it a competitive edge over other translation systems. A key advantage is its flexibility and context-awareness—it can adjust the generated content according to the given prompt. This allows for the fine-tuning of translation outputs through iterative interactions, a capability not available in traditional machine translation systems (Gao et al., 2023; Hendy et al., 2023; Kocoń et al., 2023; Lee, 2023). Gao et al. (2023) reported a significant improvement in ChatGPT's translation performance when carefully designed prompts are used in multiple rounds of interaction compared to a single-shot scenario. Moreover, ChatGPT's multi-functionality allows it to support users at various stages of the translation process, beyond merely producing a translation. such as assisting with text comprehension, generating terminology, and refining translation drafts (Azaria et al., 2023; Haleem, Javaid, & Singh, 2022; Sahari, Al-Kadi, & Ali, 2023).

Using ChatGPT for translation can benefit students in several ways. Frequent interactions with ChatGPT help students understand its strengths and limitations through hands-on experience, thereby promoting their digital and AI literacy, which is vital for their personal development in the AI era (Kornacki & Pietrzak, 2021; Muñoz-Basols et al., 2023). In addition, ChatGPT can facilitate more efficient and personalized learning by providing instant and tailored responses, enhancing the overall learning experience (Farrokhnia et al., 2023). Despite the positive findings on ChatGPT's usefulness for translation tasks, its limitations pose significant challenges for students. First, as previously mentioned, ChatGPT may produce poor-quality translations for low-resource languages (Jiao et al., 2023; Son & Kim, 2023). Second, its outputs can be biased and offensive due to biases in its training data (Azaria et al., 2023; Ray, 2023). Students must critically discern the biases and inaccuracies in the translations and information provided by ChatGPT (Muñoz-Basols et al., 2023). Third, ChatGPT cannot interpret implied meanings or cultural nuances effectively (Farrokhnia et al., 2023), often resulting in translations that lack cultural sensitivity. Furthermore, privacy concerns have been raised, as using ChatGPT for translating important documents risks violating data privacy and infringing on intellectual property rights (Azaria et al., 2023; O'Brien, 2023). A thorough understanding of the implications of using ChatGPT in the translation process is crucial for responsible use, particularly among inexperienced students (Farrokhnia et al., 2023).

However, despite the growing focus on evaluating ChatGPT's translation outputs (Jiao et al., 2023; Kocoń et al., 2023; Lee, 2023; Son & Kim, 2023), little research has been conducted on students' willingness to adopt it for translation and the factors influencing their decision. This lack of insight hinders educators from developing targeted training programmes that enable students to effectively utilize this tool while being aware of its strengths and limitations.

**2.2. The Unified Theory of Acceptance and Use of Technology (UTAUT)**

Many studies exploring the factors influencing students' use of AI tools like ChatGPT for learning are based on the Technology Acceptance Model (TAM) proposed by Davis (1989) (e.g., Liu & Ma, 2023; Ma, Wang, Li, Wang, Pang, & Wang, 2024). Despite its manageability and potential for extension (Venkatesh & Bala, 2008; Yang & Wang, 2019), TAM has been criticized for its simplicity, limited explanatory power, and lack of practical implications for technology improvement (Chuttur, 2009; Lee, Kozar, & Larsen, 2003). Another well-established framework for explaining technology acceptance is the Unified Theory of Acceptance and Use of Technology (UTAUT), formulated by Venkatesh et al. (2003) based on eight previous models and theories. Venkatesh et al. (2003) claimed that UTAUT can explain up to 70% of the variance in people's intentions to use technology. The model's strong explanatory power has been validated by many empirical studies on technology acceptance, covering a wide range of user groups, settings, products, and geographical areas (Strzelecki, 2023; Williams, Rana, & Dwivedi, 2015; Wu, Zhang, Li, & Liu, 2022).

UTAUT comprises four core constructs: performance expectancy (PE), effort expectancy (EE), social influence (SI), and facilitating conditions (FC) (Venkatesh et al., 2003). PE, EE, and SI directly influence students' behavioral intention (BI) to use the technology, while BI and FC determine use behavior (UB). In this study, PE refers to the extent to which students believe that using ChatGPT will enhance their translation performance, highlighting perceptions of the tool's usefulness and its beneficial outcomes. PE is the strongest predictor of use intention (Davis, 1989; Venkatesh et al., 2003). EE denotes the perceived ease of using ChatGPT for translation, with the notion that users are more likely to adopt a tool they find easy to use (Venkatesh et al., 2003). SI represents the degree to which students feel that important individuals in their lives, such as peers, teachers, or mentors, believe in the necessity of using ChatGPT for translation tasks. Essentially, it reflects how much students think that these significant others expect or endorse the use of ChatGPT, which can strongly affect their own intention to use the technology. In voluntary technology use, SI influences intention through internalization and identification mechanisms, where users shape their intention based on peers' and superiors' positive views (Venkatesh & Morris, 2000; Venkatesh et al., 2003; Warshaw, 1980). FC involves the belief in the presence of organizational and technical support for using ChatGPT, including compatibility with students' existing study or work styles. FC predicts actual use behavior, as adequate support and the removal of obstacles are crucial for sustained usage (Venkatesh et al., 2003). Venkatesh et al. (2012) further suggested that, in voluntary settings, varying levels of FC might directly influence behavioral intention. This study examines how these four constructs impact students' intention to use ChatGPT for translation. The UB construct in the UTAUT model, which relates to students' actual usage of ChatGPT, is not investigated. Since intention strongly predicts UB (Venkatesh et al., 2003), understanding the factors that influence students' intention to use ChatGPT for translation provides valuable insights without compromising the feasibility of this study.

Although there is limited research on students' acceptance of using ChatGPT specifically for translation, studies on higher-education students' acceptance of AI systems for general learning have generally reported a positive attitude (Duong, Bui, Pham, Vu, & Nguyen, 2023; Strzelecki, 2023; Strzelecki & ElArabawy, 2024; Tiwari, Bhat, Khan, Subramaniam, & Khan, 2023; Wu et al., 2022). For instance, Wu et al. (2022) found that EE, PE, and SI all have a significant positive influence on students' acceptance of AI-assisted learning. Strzelecki (2023) confirmed the influence of all four UTAUT predictors on students' use of ChatGPT for learning, although the effect of SI was relatively weak. Similarly, Duong et al. (2023) reported significant effects of PE and EE in similar contexts.

However, the influence of all UTAUT constructs on students' intention to use ChatGPT has not been consistently positive. While PE is widely recognized as a key predictor of use intention, the effects of EE, SI, and FC have been mixed (Alshammari & Alshammari, 2024; Romero-Rodríguez et al., 2023; Tiwari et al., 2023). For example, Romero-Rodríguez et al. (2023) found that PE positively influenced students' intentions, but EE and SI did not. They also observed that students perceive more FC as their experience with ChatGPT increases. Similarly, Tiwari et al. (2023) reported that EE was a non-significant predictor of BI, attributing this to challenges such as confusing responses and difficulties making inquiries during peak hours. While the influence of FC has generally been positive or non-significant in most studies (Alshammari & Alshammari, 2024; Strzelecki & ElArabawy, 2024), Sobaih, Elshaer, and Hasanein (2024) noted a surprising negative correlation between FC and students' intention to use ChatGPT in Saudi Arabia.

These equivocal results possibly indicate the presence of hidden moderators and underscore the need for closer investigation (Sun & Zhang, 2005). In addition, these studies have mostly focused on using ChatGPT for general learning purposes, whereas students' perceptions of its usefulness for specific professional tasks like language translation may differ (Lee et al., 2003). The present study aims to refine the investigation into students' adoption of ChatGPT to obtain more specific and relevant findings in the context of translation. Based on the UTAUT framework, we propose the following hypotheses:

H1: PE positively correlates with BI.
H2: EE positively correlates with BI.

H3: SI positively correlates with BI.

H4: FC positively correlates with BI.

**2.3. Potential Moderating Effects of Use Experience and Specialization**

Venkatesh et al. (2003) suggested that a user's experience with a tool could influence the relationships between UTAUT constructs, a theory supported by many studies (Chang, Hajiyev, & Su, 2017; Leong, Ibrahim, Dalvi-Esfahani, Shahbazi, & Nilashi, 2018; Lin, 2011; Scherer, Siddiq, & Tondeur, 2018; Yang & Wang, 2019). They defined user experience as the number of months since a user first started using the tool and discovered that with increased experience, the impact of EE and SI on BI diminishes, while the influence of FC on UB grows. This indicates that as users become more familiar with the technology, the ease of use and social influences become less significant, whereas support conditions play a more pivotal role in shaping their use intentions.

Research in machine translation has suggested that experience could moderate the influence of performance expectancy (Vieira, O'Sullivan, Zhang, & O'Hagan, 2023; Yang & Wang, 2019). As experience grows, students may develop a deeper understanding of the tool's strengths and weaknesses. Consequently, they can more effectively utilize it to enhance their translation efficiency and accuracy, leading to heightened perceptions of its performance (Yang & Wang, 2019). However, increasing usage may also unveil problems and risks associated with the tool, such as privacy concerns, which users may not be conscious of until "they experience a specific incident – e.g., when they share private information by accident – or when they notice something different or potentially suspicious in the tool's behaviour" (Vieira et al., 2023, p. 38). The extent to which these dynamics apply to students' utilization of ChatGPT as a translation aid remains unexplored. Drawing on existing insights into the moderating effect of use experience, we extend our inquiry by proposing the following hypotheses:

H5: Experience moderates the effect of PE on BI, such that the effect strengthens with more experience.

H6: Experience moderates the effect of EE on BI, such that the effect weakens with more experience.

H7: Experience moderates the effect of SI on BI, such that the effect weakens with more experience.

H8: Experience moderates the effect of FC on BI, such that the effect weakens with more experience.

In addition to use experience, the perception of ChatGPT's utility for translation may vary among students from different academic backgrounds. Non-translation students, who constitute a significant user base for machine translation tools, have received limited attention in research on these tools (Bowker, 2022; Yamada, 2020). Compared to students specializing in translation, non-translation students tend to use a narrower range of translation tools, have less awareness of their limitations, and may lack the expertise to critically evaluate information due to their limited professional training (Bowker, 2022, 2023). A small-scale survey conducted by Sahari et al. (2023) in Saudi Arabia found that translation majors preferred Google Translate over ChatGPT, while other students favoured ChatGPT. In our current study, non-translation students may also hold a more favourable attitude towards ChatGPT than translation students. However, considering that the two systems have distinct features—ChatGPT may offer more idiomatic and accurate translations in some cases, but it supports fewer languages and processes long texts more slowly (Lee, 2023)—Sahari et al. (2023) suggested that students' preferences might vary based on their specific needs.

This study will delve deeper into this issue, exploring how the objectives of students' translation practices may impact the structural relationships between the UTAUT constructs among both translation and non-translation students (refer to Fig. 1 for the adapted theoretical model). The following hypotheses are posited:

H9: The effect of PE on BI is stronger for translation students compared with non-translation students.

H10: The effect of EE on BI is weaker for translation students compared with non-translation students.

H11: The effect of SI on BI is stronger for translation students compared with non-translation students.

H12: The effect of FC on BI is weaker for translation students compared with non-translation students.

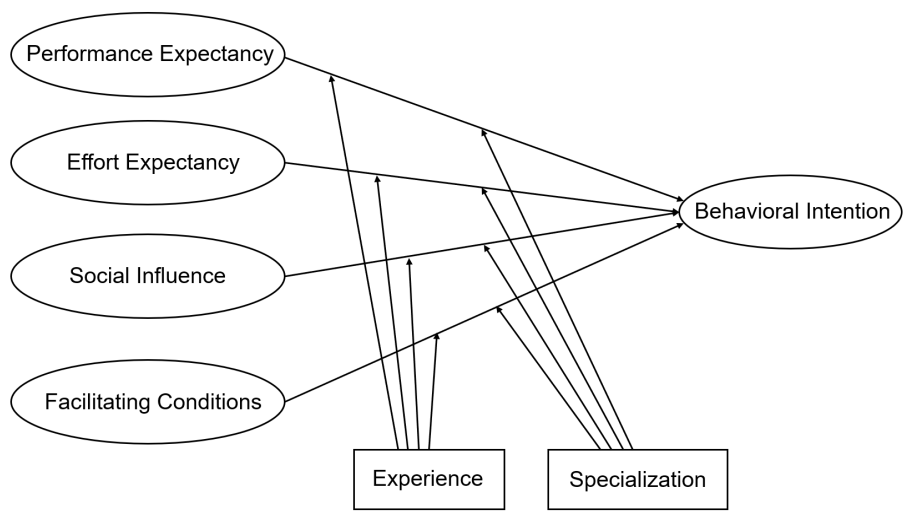

**Fig.1.** Research model.

## 3. Method
### 3.1. Sampling and Data Collection

This study adopted a cross-sectional questionnaire survey to explore the factors influencing students' adoption of ChatGPT as a translation tool. The target population comprised both translation and non-translation students currently enrolled in universities in Hong Kong. Eligibility criteria included experience with using ChatGPT for translation purposes. To determine the appropriate sample size, a power analysis was conducted using the G*Power software (Faul, Erdfelder, Lang, & Buchner, 2007; Memon, Ting, Cheah, Ramayah, Chuah, & Cham, 2020). With four predictors, two moderators (specialization and experience), and a significance level of .05, the analysis indicated a minimum sample size of 149 to detect an effect size of 0.2 with 80% power. The final sample size exceeded this requirement.

The questionnaire was administered online via Google Forms from December 2023 to February 2024. We employed convenience and snowball sampling techniques (Emerson, 2015; Wagner, 2015) to initially recruit 323 participants, consisting of 159 translation students and 164 non-translation students. Importantly, no data were missing as all fields in the questionnaire were mandatory. During the data cleaning phase, responses indicating a lack of engagement or coherence were excluded. For instance, some responses comprised repetitive entries, such as strings of repeated numbers, which suggested random input. These responses were deemed unreliable and were consequently removed from the dataset. As a result, 308 valid responses (104 male; 204 female) were retained for analysis, with 148 from translation students and 160 from non-translation students. Among the translation students, all were enrolled in English-Chinese translation programmes and were native Chinese speakers with English as their second language. Non-translation students pursued studies across various disciplines, ranging from arts and humanities to engineering and sciences. Within the sample of 308 respondents, 33.1% reported using ChatGPT 3.5, 13.3% used ChatGPT 4, and 53.6% used both versions. Detailed demographic information is presented in Table 1. Respondents were briefed on the study's purpose and confidentiality protocols at the beginning of the questionnaire. It was explicitly stated that completing the questionnaire implied consent to participate in the research. As a gesture of gratitude for their time and contribution, each respondent received a modest token of appreciation.

**Table 1**
Demographic characteristics of respondents ($N = 308$).

| Demographic Characteristic | n | % |
|---|---|---|
| Gender | | |
|     Male | 104 | 33.8 |
|     Female | 204 | 66.2 |
| Age | | |
|     18–24 | 141 | 45.8 |
|     25–30 | 141 | 45.8 |
|     ⩾31 | 26 | 8.4 |
| Education Level | | |
|     Undergraduate | 14 | 4.5 |
|     Master's | 158 | 51.3 |
|     Doctoral | 136 | 44.2 |
| Specialization | | |
|     Translation | 148 | 48.1 |
|     Non-Translation | 160 | 51.9 |
| Versions of ChatGPT Used | | |
|     GPT-3.5 | 102 | 33.1 |
|     GPT-4 | 41 | 13.3 |
|     Both | 165 | 53.6 |
| Institution | | |
|     The Hong Kong Polytechnic University | 229 | 74.4 |
|     The University of Hong Kong | 25 | 8.1 |
|     City University of Hong Kong | 23 | 7.5 |
|     The Chinese University of Hong Kong | 14 | 4.5 |
|     Hong Kong Baptist University | 8 | 2.6 |
|     Lingnan University | 4 | 1.3 |
|     The Education University of Hong Kong | 1 | 0.3 |
|     Other Hong Kong Universities | 4 | 1.3 |

### 3.2. Instruments

The questionnaire consists of three sections. To screen out ineligible respondents, it begins with a screening question: "Have you used ChatGPT for translation?" A negative response to this question results in the exclusion of the respondent from the study. The first section of the questionnaire contains items that assess the UTAUT constructs. These item statements (see Appendix) were adapted from validated scales in previous studies to suit the context of the present research (Bernabei, Colabianchi, Falegnami, & Costantino, 2023; Shroff & Keyes, 2017; Venkatesh et al., 2003; Venkatesh, Thong, & Xu, 2012). Each of the UTAUT constructs, namely PE, EE, SI, FC, and BI, is measured by four 7-point Likert scale items (1 = Strongly Disagree to 7 = Strongly Agree).

Section 2 collects data on the frequency of students' usage of various translation tools and resources (Bowker, 2002; Gough, 2017), including MT systems, search engines, dictionaries, corpora, grammar checkers, computer-aided translation (CAT) tools, professional translation services, and ChatGPT. Students rate their usage frequency of these resources on a 5-point scale (1 = Never to 5 = Always). This aims to gain insights into the role of ChatGPT in students' translation toolboxes, thereby helping address RQ2. Additionally, this section gathers data on students' purposes for engaging in translation activities and their experience using ChatGPT as a translation tool. Section 3 collects demographic information, including age, gender, level of education, institution, and area of study.

Prior to the large-scale distribution of the questionnaire, we sought input from three experts in the field of translation studies regarding the questionnaire items. We incorporated their feedback and made necessary revisions to enhance clarity and relevance. Subsequently, the questionnaire underwent a pilot phase involving 10 translation students and 10 non-translation students from the target population. Their feedback was gathered to identify any items that were perceived as ambiguous or incomprehensible. Any such items were then revised or removed accordingly to ensure the questionnaire's effectiveness.

### 3.3. Data Analysis

To investigate the moderating influence of experience on the relationships between the UTAUT constructs, participants were categorized into two groups based on their level of experience using ChatGPT. Previous studies have often measured experience as the duration since the initial adoption of a tool (Awwad & Al-Majali, 2015; Romero-Rodríguez et al., 2023; Venkatesh et al., 2012). While this approach is convenient, its accuracy may be questioned as it overlooks usage frequency. In our study, we considered both the number of months since respondents first used ChatGPT and their self-reported frequency of use for translation tasks. Usage frequency was assessed across three levels: 3 = often (daily/weekly), 2 = sometimes (monthly), and 1 = rarely. Subsequently, we multiplied these two values and categorized the 308 respondents into high-experience (n = 154) and low-experience (n = 154) groups based on the median of the interaction term (Iacobucci, Posavac, Kardes, Schneider, & Popovich, 2015).

Data analysis was conducted with SPSS 26 and AMOS 28. Initially, we performed a confirmatory factor analysis (CFA) to assess distribution normality, construct reliability, and validity, as well as model fit. Construct reliability was evaluated using composite reliability and Cronbach's alpha, which indicate the internal consistency of items within each construct (Taber, 2017). Construct validity was examined through convergent and discriminant validity (Hair, Babin, & Krey, 2017). Subsequently, we tested the structural model using maximum likelihood estimation in covariance-based structural equation modeling (CB-SEM), a widely-used method for examining relationships between latent variables (Kline, 2023). Multi-group analysis was conducted to explore the moderating effects of use experience and specialization (Awang, 2012; Byrne, 2004).

**4. Results**
**4.1. Assessment of Measurement Model and Measurement Invariance**
We first checked the normality of each measurement item. The skewness ranged from -1.36 to 0.29, and the kurtosis ranged from -0.95 to 2.32. Their absolute values were smaller than 2.0 and 7.0, respectively, indicating acceptable normality (Finney & DiStefano, 2013). The average ratings for each item were above the midpoint, suggesting that respondents generally held positive views toward using ChatGPT for translation. Next, we examined construct reliability and validity (see Table 2). The values of Cronbach's alpha all exceeded 0.7, and the factor loadings of all items were significant, demonstrating high internal consistency of the measurement scale (Taber, 2017). Composite reliability values were also greater than 0.7, and the average variance extracted (AVE) values were above 0.5, thus establishing good reliability and convergent validity (Fornell & Larcker, 1981; Hair et al., 2017).

**Table 2**

Confirmatory factor analysis and reliability assessment ($N = 308$).

| Constructs | Items | Mean | SD | Standardized Factor Loading | Cronbach's alpha | CR | AVE |
|---|---|---|---|---|---|---|---|
| Performance Expectancy (PE) | PE1 | 5.75 | 1.19 | 0.85 | 0.90 | 0.90 | 0.69 |
| | PE2 | 5.83 | 1.30 | 0.83 | | | |
| | PE3 | 5.69 | 1.24 | 0.85 | | | |
| | PE4 | 5.62 | 1.26 | 0.78 | | | |
| Effort Expectancy (EE) | EE1 | 5.77 | 1.26 | 0.76 | 0.85 | 0.85 | 0.59 |
| | EE2 | 5.37 | 1.22 | 0.73 | | | |
| | EE3 | 5.60 | 1.29 | 0.83 | | | |
| | EE4 | 5.34 | 1.32 | 0.75 | | | |
| Social Influence (SI) | SI1 | 4.94 | 1.62 | 0.88 | 0.82 | 0.83 | 0.63 |
| | SI2 | 4.77 | 1.63 | 0.86 | | | |
| | SI3 | 4.92 | 1.62 | 0.61 | | | |
| Facilitating Conditions (FC) | FC1 | 5.51 | 1.44 | 0.70 | 0.76 | 0.77 | 0.53 |
| | FC2 | 5.36 | 1.34 | 0.81 | | | |
| | FC3 | 5.33 | 1.37 | 0.66 | | | |
| Behavioral Intention (BI) | BI1 | 5.01 | 1.59 | 0.84 | 0.89 | 0.90 | 0.68 |
| | BI2 | 5.21 | 1.47 | 0.83 | | | |
| | BI3 | 4.85 | 1.65 | 0.82 | | | |
| | BI4 | 5.59 | 1.56 | 0.82 | | | |

*Note*. CR: Composite Reliability, AVE: Average Variance Extracted.
Items SI4 and FC4 were removed due to low factor loadings (below 0.5), indicating that they explained little of their respective constructs and compromised composite reliability and model fit. The remaining items were adequate to ensure statistical identification of SI and FC (Awang, 2012; Hair et al., 2017).

Discriminant validity was assessed by calculating the heterotrait–monotrait (HTMT) ratio of correlations between constructs (Hensseler et al., 2015) and comparing the square roots of AVEs with the correlations among constructs (Fornell & Larcker, 1981). All HTMT ratios were below the threshold of 0.85 (see Table 3), and the square roots of the AVEs for each construct exceeded their respective correlations with other constructs (Table 4). These findings collectively indicate strong discriminant validity of the measurement model.

**Table 3**

HTMT ratio of correlations.

|  | PE | EE | SI | FC | BI |
|---|---|---|---|---|---|
| Performance Expectancy (PE) |  |  |  |  |  |
| Effort Expectancy (EE) | 0.74 |  |  |  |  |
| Social Influence (SI) | 0.26 | 0.22 |  |  |  |
| Facilitating Conditions (FC) | 0.56 | 0.68 | 0.33 |  |  |
| Behavioral Intention (BI) | 0.74 | 0.61 | 0.30 | 0.58 |  |

**Table 4**

Correlation matrix and discriminant validity.

|  | PE | EE | SI | FC | BI |
|---|---|---|---|---|---|
| PE | 0.83 |  |  |  |  |
| EE | 0.74 | 0.77 |  |  |  |
| SI | 0.26 | 0.22 | 0.79 |  |  |
| FC | 0.56 | 0.68 | 0.33 | 0.73 |  |
| BI | 0.73 | 0.61 | 0.30 | 0.58 | 0.83 |

*Note*. The numbers along the diagonal represent the square roots of AVEs. Other numbers in the table are the correlation coefficients between the UTAUT constructs.

We then evaluated the fit of the measurement model using the Comparative Fit Index (CFI, with values above 0.90 indicating good fit), the Tucker-Lewis Index (TLI, > 0.90), the Root Mean Square Error of Approximation (RMSEA, < 0.08), and the Standardized Root Mean Square Residual (SRMR, < 0.08) (Hair et al., 2017; Hu & Bentler, 1999; Kline, 2023). The model exhibited good fit to the data, $\chi^2(125) = 286.81$, $\chi^2/df = 2.30$, CFI = 0.95, TLI = 0.94, RMSEA = 0.065, 90% CI [0.055, 0.075], SRMR = 0.05.

Prior to conducting the multigroup SEM, measurement invariance across groups was assessed through multigroup CFA, following the guidelines of Byrne (2004, 2008). As shown in Table 5, the CFA model with experience as the grouping variable demonstrated good fit, suggesting configural equivalence across students with varying levels of ChatGPT use experience (Byrne, 2004; Hair et al., 2017). To investigate whether the items were interpreted similarly across groups, we constrained factor loadings to be equal across the two groups and evaluated model fit. Results indicated no significant decrease in model fit ($\Delta\chi^2 = 20.03$, $\Delta df = 16$, $p = .219$), supporting metric invariance (Byrne, 2004). Similarly, the unconstrained model with specialization as the grouping variable displayed acceptable model fit, indicating configural invariance across translation and non-translation students. When factor loadings were constrained across the two groups, there was also no significant decrease in model fit, $\Delta\chi^2 = 12.23$, $\Delta df = 13$, $p = .509$.

**Table 5**
Comparison of fit between unconstrained and constrained measurement models.

| Models | $\chi^2(df)$ | CFI | TLI | RMSEA | SRMR |
|---|---|---|---|---|---|
| Baseline | 286.81(125) | 0.95 | 0.94 | 0.065 [0.055, 0.075] | 0.05 |
| Experience-grouped | 443.86(250) | 0.94 | 0.93 | 0.05 [0.043, 0.058] | 0.05 |
| Constrained experience-grouped | 462.64(263) | 0.94 | 0.93 | 0.05 [0.042, 0.057] | 0.06 |
| Specialization-grouped | 480.12(250) | 0.93 | 0.91 | 0.055 [0.047, 0.062] | 0.06 |
| Constrained specialization-grouped | 492.34(263) | 0.93 | 0.91 | 0.053 [0.046, 0.061] | 0.06 |

*Note.*
Baseline: the baseline model with the entire sample.
Experience-grouped: unconstrained model with use experience as the grouping variable.
Constrained experience-grouped: model with factor loadings constrained to equality across groups with high and low experience levels.
Specialization-grouped: unconstrained model with students' specialization as the grouping variable.
Constrained specialization-grouped: model with factor loadings constrained to equality across translation and non-translation student groups.

**4.2. Structural Model**

Results of the SEM indicated a good fit of the structural model, $\chi^2(125) = 286.81$, $\chi^2/df = 2.30$, CFI = 0.95, TLI = 0.94, RMSEA = 0.065, 90% CI [0.055, 0.075], SRMR = 0.05 (Hair et al., 2017; Hu & Bentler, 1999; Kline, 2023). Table 6 presents the path coefficient estimates of the structural model ($N = 308$). Specifically, BI was positively influenced by PE ($\beta = 0.59$, $p < .001$) and FC ($\beta = 0.22$, $p = .006$), but was non-significantly influenced by EE ($\beta = 0.01$, $p = .894$) and SI ($\beta = 0.07$, $p = .183$). The four predictors together explained 58.9% of the variance of BI. Based on these results, H1 and H4 were accepted, while H2 and H3 were rejected.

**Table 6**
Hypothesis testing with the entire sample.

| Paths | Standardized Regression Coefficients | t-values | p-values | Result |
|---|---|---|---|---|
| H1: PE→BI | 0.59 | 6.90 | < .001*** | Supported |
| H2: EE→BI | 0.01 | 0.13 | .894 | Rejected |
| H3: SI→BI | 0.07 | 0.18 | .183 | Rejected |
| H4: FC→BI | 0.22 | 2.76 | .006** | Supported |

*Note.* *** significant at the 0.001 level, ** significant at the 0.01 level.

### 4.2.1. Multigroup Analysis by Experience

The unconstrained structural model with experience as the grouping variable showed good fit, $\chi^2(250) = 443.86$, $\chi^2/df = 1.78$, CFI = 0.94, TLI = 0.93, RMSEA = 0.05, 90% CI [0.043, 0.06], SRMR = 0.058. When equality constraints were applied to the structural weights across the high-experience and low-experience groups, we did not observe a significant decrease in model fit, $\Delta\chi^2 = 24.17$, $\Delta df = 17$, $p = .115$. However, the structural relationships exhibited different patterns when examined separately for each group. As illustrated in Fig. 2, PE was a key predictor of BI for both high-experience and low-experience students, but the correlation between PE and BI was slightly stronger for experienced students ($\beta = 0.63$, $t = 5.60$, $p < .001$) than for less-experienced students ($\beta = 0.42$, $t = 2.92$, $p = .004$). Therefore, H5 was confirmed. Effort expectancy was not a significant predictor for BI in either group, and H6 was not supported. In addition, social influence significantly correlated with BI only for less-experienced students ($\beta = 0.25$, $t = 2.89$, $p = .004$), thus confirming H7. The correlation between facilitating conditions and BI was non-significant for both groups, so H8 was rejected. Overall, the model accounted for 56.9% of experienced students' and 60% of less-experienced students' intention to use ChatGPT for translation purposes.

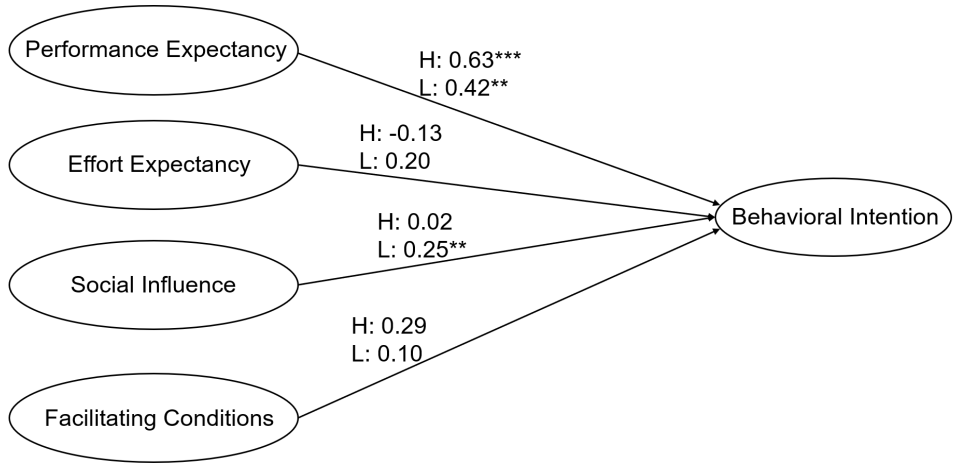

**Fig. 2.** Standardized path coefficients for high-experience and low-experience students.

*Note.* ** *p* ≤ .01, *** *p* ≤ .001. H: high-experience group, L: low-experience group.

### 4.2.2. Multigroup Analysis by Specialization

With students' specialization as the grouping variable, the unconstrained structural model demonstrated good fit to the data, $\chi^2(250) = 480.12$, $\chi^2/df = 1.92$, CFI = 0.93, TLI = 0.91, RMSEA = 0.055, 90% CI [0.047, 0.062], SRMR = 0.06. We then compared the path coefficients between the two groups (see Fig. 3). The correlation between PE and BI was slightly stronger for translation majors ($\beta = 0.70$, $t = 5.12$, $p < .001$) compared with non-translation majors ($\beta = 0.56$, $t = 4.74$, $p < .001$), thus supporting H9. EE and SI did not significantly correlate with BI for either group. Therefore, H10 and H11 were rejected. FC significantly predicted BI only for non-translation students ($\beta = 0.29$, $t = 2.66$, $p = .008$) and not for translation students' BI ($\beta = 0.06$, $t = 0.44$, $p = .663$), which supported H12. Overall, the model explained 58.1% of the variance in BI among translation students and 61.1% among non-translation students.

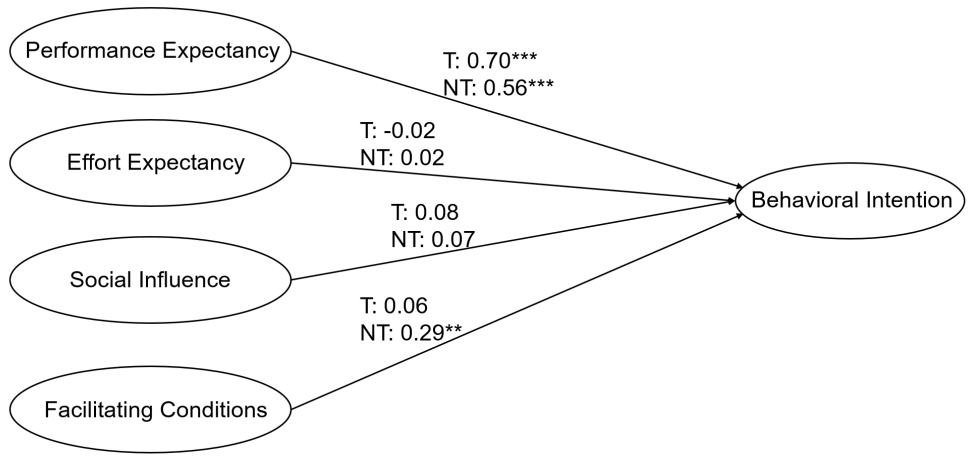

**Fig. 3.** Standardized path coefficients for translation and non-translation students.
*Note*. ** *p* ≤ .01, *** *p* ≤ .001. T: translation students, NT: non-translation students.

### 4.3. Student Ratings of Translation Tool Usage Frequency

According to the frequency ratings shown in Fig. 4, MT tools were generally the most frequently used translation aids by students, with their usage frequency exceeding that of ChatGPT. This preference was statistically supported by a Wilcoxon signed-rank test, which indicated that the difference in usage frequency between MT tools and ChatGPT was significant, with *z* = -4.06, *p* < .001 for translation students and *z* = -2.59, *p* = .009 for non-translation students. Despite this, ChatGPT remained among the top three translation tools preferred by students. Furthermore, there was no significant difference in ChatGPT usage frequency for translation purposes between translation and non-translation students.

To compare the usage frequency of translation tools between translation and non-translation students, a Mann-Whitney U test was conducted. The results showed that translation students reported more frequent usage of search engines (*z* = -4.59, *p* < .001), dictionaries (*z* = -7.93, *p* < .001), and corpora (*z* = -4.24, *p* < .001) during their translation process compared to non-translation students. This suggests that translation majors tend to utilize a broader range of resources when translating compared to non-translation majors.

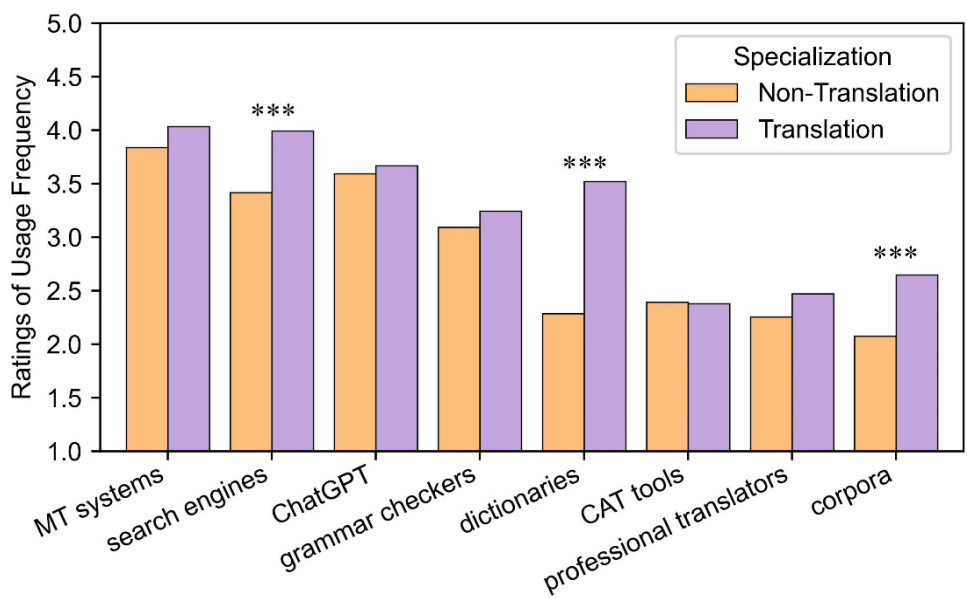

**Fig. 4.** Usage frequency of translation tools and resources.

*Note.* *** *p* ≤ .001.

### 4.4. Purposes of Students' Translation Practices

To better understand students' acceptance of ChatGPT as a translation tool, we compared the common purposes of translation activities between translation and non-translation students (refer to Fig. 5). Our findings indicate that both groups primarily engage in translation for reading comprehension, writing, and improving language proficiency, while translating solely for translation tasks is less common and occurs only occasionally. A Mann-Whitney U test revealed significant differences between the predominant purposes of translation for the two groups. Non-translation students translate more frequently for reading comprehension ($z$ = -1.95, $p$ = .05) and writing tasks ($z$ = -2.21, $p$ = .027), whereas translation students engage in translation more often to improve language proficiency ($z$ = -2.45, $p$ = .014), complete translation tasks ($z$ = -3.34, $p$ = .001), and for entertainment ($z$ = -3.88, $p$ < .001). It appears that for non-translation students, translation serves as a means to other ends rather than being an end in itself.

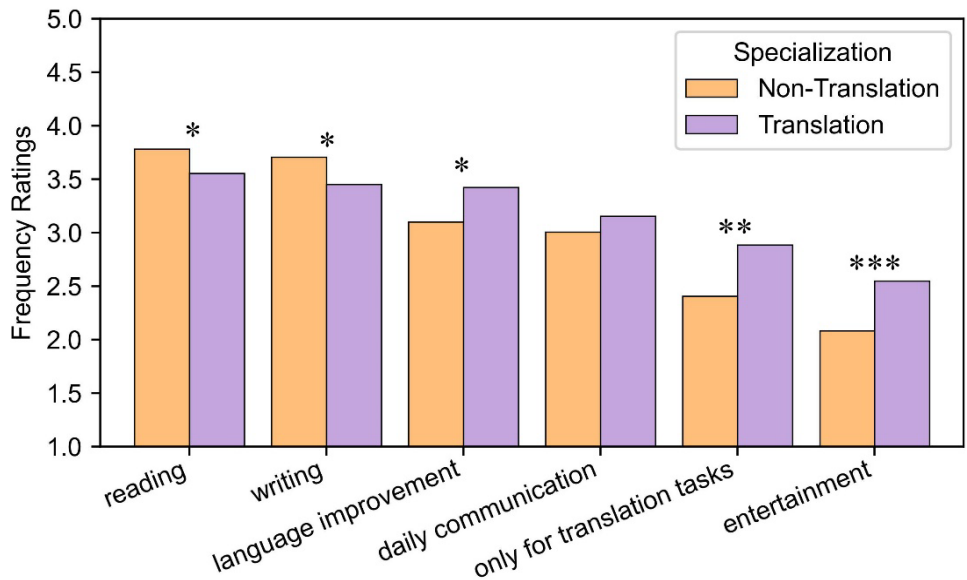

**Fig. 5.** Purposes of translation practices.
*Note.* * $p \leq .05$, ** $p \leq .01$, *** $p \leq .001$.

## 5. Discussion

This study utilized the UTAUT model to investigate the factors influencing university students' intention to use ChatGPT as a translation tool in Hong Kong. It also examined the moderating effects of students' ChatGPT use experience and specialization through multigroup structural equation modeling. The survey findings revealed that, although students generally had positive perceptions regarding using ChatGPT for translation, machine translation systems remained their most commonly used translation tools. Structural equation modeling conducted on the entire student sample partially supported the UTAUT model.

Firstly, PE emerged as the strongest determinant of students' intention to use ChatGPT for translation. This suggests that students are primarily motivated by the perceived benefits and usefulness of the technology in enhancing their translation performance. This finding is consistent with numerous studies on ChatGPT acceptance (Alshammari & Alshammari, 2024; Romero-Rodríguez et al., 2023; Strzelecki, 2023; Strzelecki & ElArabawy, 2024) and is aligned with Venkatesh et al.'s (2003) conceptualization of the UTAUT model. The model posits that individuals are more likely to adopt a technology when they believe it will help them achieve better outcomes. As translation training programmes consider integrating ChatGPT and other AI-based tools into their curricula, it is essential to highlight and demonstrate both the benefits and risks these technologies might pose to students' learning outcomes and linguistic skills. Moreover, the significance of PE in shaping students' intentions to use ChatGPT for translation underscores the need for ongoing research and development to optimize the tool's translation-related capabilities and align them with the specific needs and expectations of students. Such customization may help reinforce students' perceptions of ChatGPT's usefulness and encourage wider adoption in educational settings.

Furthermore, consistent with the findings of Alshammari and Alshammari (2024), Romero-Rodríguez et al. (2023), and Strzelecki (2023), apart from PE, students' intention to use ChatGPT for translation was also significantly influenced by their perceptions of FC. According to Venkatesh et al. (2003), having the necessary resources, support, and infrastructure positively affects one's decision to adopt new technologies. When students perceive that they have easy access to ChatGPT, along with adequate technical support and guidance on how to use it effectively, and that it is compatible with other translation tools they use, they are more likely to embrace it for translation. To promote the adoption of ChatGPT for translation, interested parties should focus on creating a supportive environment that enables students to seamlessly integrate the tool into their translation workflows.

Interestingly, EE and SI did not significantly impact students' intention to use ChatGPT for translation. This result diverges from the original UTAUT model (Venkatesh et al., 2003) and some studies on ChatGPT acceptance (Strzelecki, 2023; Strzelecki & ElArabawy, 2024; Sobaih, Elshaer, & Hasanein, 2024), but aligns with the findings of Alshammari and Alshammari (2024) and Romero-Rodríguez et al. (2023). It appears that students' perceptions of the ease associated with using ChatGPT for translation and the influence of their social environment are less critical in shaping their use intention compared with PE and FC. Notably, the influence of EE was not observed for all sub-samples of students. The non-significant influence of EE may be attributed to the digital adeptness of students and the user-friendliness of ChatGPT's interface. As the majority of our survey respondents were below 30 years old, they are likely to be digital natives who are accustomed to learning and using new technologies (Helsper & Eynon, 2010). For these students, the potential barriers to using ChatGPT for translation may not constitute challenges that hinder their adoption of it. Moreover, as a chatbot, ChatGPT's interface features conversational interactions similar to those in chat applications. The user-friendly nature of the interface may further diminish the impact of EE on students' intention to use it for translation tasks.

Importantly, the relative importance of the other three UTAUT constructs varied across student populations with different levels of ChatGPT use experience and from different disciplinary backgrounds. We found that for students with comparatively limited experience using ChatGPT, their intention to use it for translation was driven by PE and SI. However, for more experienced students, the impact of SI became non-significant, while the influence of PE grew more salient. The findings resonate with the findings of Venkatesh et al. (2003) and Venkatesh and Morris (2000), which can help explain the differences in factors influencing students' intention to use ChatGPT for translation based on their experience level. Novice students, who have limited interaction with ChatGPT, tend to rely heavily on the opinions and influences of important individuals and credible sources when forming their intention to use the technology. This results in a stronger perceived social influence for these less-experienced students. As students gain more hands-on experience with ChatGPT, they develop their own evaluations of its efficacy for translation purposes, focusing more on whether the tool meets their translation needs. At this stage, students' independent judgment outweighs social influence, leading to the non-significant role of SI and the more decisive role of PE in determining their intention to use ChatGPT. These findings suggest that leveraging the influence of teachers, peers, and other important and trusted sources could be an effective way to encourage or discourage initial adoption of ChatGPT as a translation tool among novice students, but the perceived usefulness and benefits are the primary driver of sustained use.

In terms of the moderating effect of specialization on students' acceptance and use of ChatGPT, although the importance of ChatGPT in translation and non-translation students' translation toolbox did not differ significantly (as illustrated in Fig. 4), the two student populations' intention to use ChatGPT for translation was shaped by different factors. While the impacts of EE and SI were non-significant for both groups, translation students' use intention was predominantly driven by PE, and for students in other disciplines, both PE and FC played important roles in influencing use intention. These results suggest that translation students prioritize performance benefits when deciding whether to use the tool, while institutional or technical support did not matter to them. By comparison, non-translation students consider both performance outcomes and the supporting environment.

This difference in focus could be attributed to their diverse translation needs, levels of translation expertise, and tech-savviness. As depicted in Fig. 5, translation students' practices involved more professional translation tasks and intentions to improve language proficiency, while non-translation students more frequently translated for other purposes, such as facilitating reading comprehension and writing. Driven by their professional and academic goals, translation students may prioritize the performance benefits that ChatGPT offers in enhancing their translation quality and efficiency. They may be more willing to invest time and effort in mastering the technology, regardless of the availability of extensive institutional support. However, non-translation students use ChatGPT for more diverse and less specialized purposes. They may place greater value on the ease of use and availability of support, as they might lack the same strong intrinsic motivation to invest in mastering the tool.

Moreover, given the limited translation training and expertise of non-translation students, they might lack the discernment to fully evaluate ChatGPT's performance (Bowker, 2023). In contrast, translation students typically integrate a wider array of technological aids into their work (see Fig. 4). When utilizing ChatGPT for translation, they likely explore a broader range of its features, including eliciting subject knowledge, preparing terminology, and proofreading translation outputs. These explorations, combined with their translation expertise and sharper discernment, enable them to more keenly perceive the advantages and limitations of ChatGPT, aligning directly with their academic and professional objectives. Consequently, translation students may exhibit stronger perceptions of the performance outcomes associated with using ChatGPT for translation tasks. Hence, the impact of performance expectancy on translation students' intention to use ChatGPT was more pronounced compared to non-translation students.

Finally, the tech-savviness of translation students may also weaken the influence of facilitating conditions on their intention to use ChatGPT. Due to their frequent utilization of various resources during translation, spanning from convenient online machine translation platforms to complex CAT tools, and the incorporation of translation technology into their curricula (Chan, Chow, & Wong, 2014; Liu, Kwok, Liu, & Cheung, 2022), translation students are likely more adept at learning and adapting to new technologies. This tech-savviness may diminish their reliance on institutional or technical support when utilizing ChatGPT for translation, as they are more capable of independently navigating the technology. In contrast, non-translation students, who may have less exposure to translation technologies and limited integration of such tools in their curricula, may place greater importance on the availability of support and resources when deciding to use ChatGPT for translation.

## 6. Conclusion

This study holds both theoretical and practical significance. Theoretically, it contributes to validating the UTAUT model in the specialized academic domain by focusing on the AI chatbot ChatGPT. By examining student acceptance of ChatGPT specifically for translation tasks, it offers a context-specific perspective, complementing previous research in educational settings. This study enhances our comprehension of the factors influencing students' adoption of ChatGPT as a translation tool and provides empirical groundwork for future investigations into AI-assisted translation training. Furthermore, it sheds light on potential moderators by exploring how domain expertise and ChatGPT use experience may impact the relationships between UTAUT constructs within this context. The implications of these findings extend to both translation curriculum designers and developers of translation tools. They underscore the pivotal role of performance expectancy in students' decisions to utilize ChatGPT for translation activities across diverse disciplines and levels of experience. This highlights the importance of educating students about the potential benefits and risks associated with using ChatGPT as a translation aid, empowering them to make informed decisions about its usage. Moreover, the significance of facilitating conditions for non-translation students warrants further investigation and attention. By addressing perspectives from non-translation students and uncovering the factors influencing their adoption of ChatGPT as a translation tool, this study contributes to the ongoing efforts to enhance non-translation students' machine translation literacy. Understanding the correlation between students' translation needs and their motivations for using ChatGPT could offer valuable insights for course designers and translation tool developers.

It should be noted that this study was conducted in Hong Kong, and the findings cannot be generalized to students in other geographical regions. Future research could explore the roles of various moderators, including geographical location, age, and gender, to provide a more comprehensive understanding of students' acceptance of AI tools. Additionally, this study is limited by considering only two levels of use experience, and the self-reported data may contain inaccuracies. A more detailed exploration of user experience levels could provide a more nuanced understanding of its moderating impacts. Furthermore, the quantitative data in this study have not been corroborated by qualitative evidence. Future work could focus on the discrepancies between the findings of this study and those of previous research to deepen the understanding of the relationships among the UTAUT constructs.

**Data Availability:** Data concerning the study are publicly available on Open Science Framework (https://osf.io/yh8p3/).

**Appendix**

*UTAUT Questionnaire Items*

| Constructs | Items | Source |
|---|---|---|
| Performance expectancy | I find ChatGPT useful for translation. Using ChatGPT enables me to accomplish translation tasks more quickly. Using ChatGPT increases my translation productivity. Using ChatGPT increases my chances of achieving my goals. | Venkatesh et al., 2003 |
| Effort expectancy | Learning to use ChatGPT is easy for me. My interaction with ChatGPT is clear and understandable. I find ChatGPT easy to use for translation. It is easy for me to become skillful at using ChatGPT for translation. | Venkatesh et al., 2003 |
| Social influence | I use ChatGPT for translation because people around me have mentioned it. I use ChatGPT for translation because people around me use it. I use ChatGPT for translation because I heard about it from lectures/social networks/newspapers. I use ChatGPT for translation to stay updated. (Removed) | Bernabei et al., 2023 |
| Facilitating conditions | I have the resources necessary to use ChatGPT for translation (e.g., access permission). I have the knowledge necessary to use ChatGPT for translation (e.g., knowledge of ChatGPT's capabilities). ChatGPT is compatible with other translation tools and resources I use. I can get help from others when I have difficulties using ChatGPT for translation. (Removed) | Venkatesh et al., 2003 |
| Behavioral intention to use | I intend to use ChatGPT frequently for my translation tasks. I intend to increase my use of ChatGPT for translation tasks. Whenever possible, I intend to use ChatGPT for translation as often as needed. I expect to continue using ChatGPT as a translation tool in the future. | Shroff & Keyes, 2017 |